\documentclass[letterpaper,12pt]{article}

\usepackage{comment}
\usepackage[T1]{fontenc}
\usepackage{graphicx}
\usepackage{todonotes}
\usepackage{booktabs}
\usepackage{caption} 
\usepackage[font=small]{caption}
\usepackage[utf8]{inputenc}
\usepackage{mathtools}
\usepackage{ upgreek }
\usepackage{tikz}
\usepackage{listings}
\usepackage{subfig}

\usepackage{hyperref}

\newcommand{\removeStuff}[1]{}
\newcounter{incFigCounter}
\stepcounter{incFigCounter}
\newcommand{\incFig}[1]{%
\includegraphics{\arabic{incFigCounter}.pdf}%
\stepcounter{incFigCounter}%
}

\newsavebox{\measurebox}
\newcommand{\charsg}{\mathcal{S}(\CN)}

\newcommand{\cay}{\mathrm{Cay}}
\newcommand{\pcay}{\mathrm{PCay}}
\newcommand{\scay}{\mathrm{SCay}}
\newcommand{\orbit}{\mathcal{O}}
\newcommand{\CN}{\mathrm{CN}}
\newcommand{\TCA}{\text{TCA-}\square}
\newcommand{\sTCA}{\text{TCA-}\tetrahedron}

\newcommand{\figname}{Fig.~}
\newcommand{\defname}{Def.~}

\newcommand{\tetrahedron}{
  \mathchoice
    {\includegraphics[height=1.5ex]{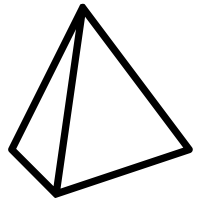}} %
    {\includegraphics[height=1.5ex]{tetrahedron}} %
    {\includegraphics[height=.8ex]{tetrahedron}} %
    {\includegraphics[height=.5ex]{tetrahedron}} %
}

\usepackage{amsthm}
\newtheorem{observation}{Observation}
\newtheorem{definition}{Definition}
\usepackage{chemformula}

\newcommand{\myTitle}{Cayley Graphs of Semigroups Applied to Atom Tracking in Chemistry}

    \usepackage{abstract} %

    \usepackage{titlesec} %

    \usepackage{authblk} %

	\usepackage{datetime} %
    \newdateformat{usvardate}{
  	\monthname[\THEMONTH] \ordinal{DAY}, \THEYEAR}

  	\usepackage[bottom]{footmisc} %

    \usepackage{fancyhdr}
    \fancyheadoffset{0cm}
    \providecommand{\keywords}[1]{\textbf{\textit{Keywords---}} #1}

    \usepackage{amssymb}
    \usepackage{amsthm}
    \usepackage{newtxmath}

    \usepackage{microtype} %
    \usepackage[utf8]{inputenc}

	\usepackage{setspace} %
    \usepackage{lineno}

	\usepackage[top=1.5in, bottom=1.5in, left=1in, right=1in]{geometry}

    \usepackage{graphicx} %
    \usepackage{float} %

    \usepackage{printlen}

    \usepackage{longtable} %
    \usepackage{tabularx} %
    \newcolumntype{L}[1]{>{\raggedright\arraybackslash}p{#1}}
    \newcolumntype{C}[1]{>{\centering\arraybackslash}p{#1}}
    \newcolumntype{R}[1]{>{\raggedleft\arraybackslash}p{#1}}
    \usepackage{relsize} %

	\usepackage{hyperref} %

        \usepackage[
        style=authoryear,
	 eprint=false,
	 dashed=false,
	 maxbibnames=2,
	 minbibnames=2,
	 doi=false,
         isbn=false,
	 backend=bibtex8
]{biblatex}
\DeclareNameAlias{sortname}{last-first}
\bibliography{refs}
\def\cite{\parencite}

\begin{document}

    \thispagestyle{empty} %

      \title{\vspace{-15mm}\fontsize{21pt}{10pt}\selectfont\textbf{\myTitle }}

    	\date{Submitted: \usvardate\today}

    \makeatletter
    \renewcommand{\@maketitle}{
    \newpage
     \null
     \vskip 2em%
     \begin{center}%
      {\LARGE \@title \par}%
     \end{center}%
     \par} \makeatother

    \maketitle %

    \begin{abstract}
      \noindent 
	  While atom tracking with isotope-labeled compounds is an essential
	  and sophisticated wet-lab tool in order to, e.g., illuminate
	  reaction mechanisms, there exists only a limited amount of formal
	  methods to approach the problem. Specifically when large
	  \mbox{(bio-)chemical} networks are considered where reactions are
	  stereo-specific, rigorous techniques are inevitable. We
	  present an approach using the right Cayley graph of a monoid in
	  order to track atoms concurrently through sequences of reactions and
	  predict their potential location in product molecules. This can not
	  only be used to systematically build hypothesis or reject reaction
	  mechanisms (we will use the ANRORC mechanism ``Addition of the Nucleophile,
	  Ring Opening, and Ring Closure'' as an example), but also to infer
	  naturally occurring subsystems of (bio-)chemical systems. Our results
	  include the analysis of the carbon traces within the TCA
	  cycle and infer subsystems based on projections of the right Cayley
	  graph onto a set of relevant atoms.
    \end{abstract}

    \setlength\parindent{.45in} \keywords{Computational Biology, Graph
      Transformations, Double Pushout, Chemical Reaction Networks, Algorithmic Cheminformatics}

    \doublespacing

	\section{Introduction}
	Traditionally, atom tracking is used in chemistry to understand the
	underlying reactions and interactions behind some chemical or
	biological system. In practice, atoms are usually tracked using
	isotopic labeling experiments. In a typical isotopic labeling
	experiment, one or several atoms of some educt molecule of the
	chemical system we wish to examine are replaced by an isotopic
	equivalent (e.g. $^{12}$C is replaced with $^{13}$C). These
	compounds are then introduced to the system of interest, and the
	resulting product compounds are examined, e.g. by mass
	spectrometry \cite{isotope-ms} or nuclear magnetic resonance
	\cite{isotope-nmr}. By determining the positions of the isotopes in the
	product compounds, information about the underlying reactions
	might then be derived. 
	From a theoretical perspective, characterizing a formal framework
	to track atoms through reactions is an important step to
	understand the possible behaviors of a chemical or biological
	system. In this contribution, we introduce such a framework
	based on concepts rooted in semigroup theory.
	Semigroup theory can be used as a tool 
	to analyze biological systems such as metabolic and gene regulatory
	networks \cite{nehaniv2015symmetry,egri2008hierarchical}. 
	In particular, Krohn-Rhodes theory \cite{rhodes2009applications}
	was used to analyze biological systems 
	by decomposing a semigroup into simpler components.
	The networks are modeled as state automatas (or ensembles of
	automatas), and their characteristic semigroup, i.e., the semigroup
	that characterizes the transition function of the automata
	\cite{algebraic-automata}, is then decomposed using Krohn-Rhodes
	decompositions or, if not computationally feasible, the holonomy
	decomposition variant \cite{egri2015computational}.  The result is a set of
	symmetric natural subsystems and an associated hierarchy between them,
	that can then 
	be used to reason about the system.
	In \cite{IsotopeLabel2019} algebraic structures were employed for
	modeling atom tracking: graph transformation rules are iteratively
	applied to sets of undirected graphs (molecules) in order to generate
	the hyper-edges (the chemical reactions) of a directed hypergraph (the
	chemical reactions network)
	\cite{andersen2016software,chemicalmotifs}. A semigroup is
	defined by using the (partial) transformations that naturally arise from
	modeling chemical reactions as graph transformations. Utilizing this particular
	semigroup so-called pathway tables can be constructed, detailing the
	orbit of single atoms through different pathways to help with the
	design of isotopic labeling experiments.

	In this work, we show that we can gain a deeper understanding of
	the analyzed system by considering how atoms move in relation to each
	other. To this end, we briefly introduce useful terminology in 
	Section \ref{sec:preliminaries}, found in graph
	transformation theory as well as semigroup theory.  
	In Section \ref{sec:chemical-networks-and-algebraic-structures} we show how
	the possible trajectories of a subset
	of atoms can be intuitively represented as the (right) Cayley graph
	\cite{denes1966connections} of the associated semigroup of a chemical network.  
	Moreover, we define natural subsystems of a chemical network in terms of reversible
	atom configurations and show how they naturally relate to the strongly
	connected components of the corresponding Cayley graph.
	We show the usefulness of our approach in Section \ref{sec:results}
	by using the constructions defined in Section \ref{sec:chemical-networks-and-algebraic-structures} to differentiate chemical pathways, based on the atom trajectories derived from each pathway.
	We then show how the Cayley graph additionally provides a
	natural handle for the analysis of cyclic chemical systems such as the
	TCA cycle \cite{biochem-textbook}.

	\section{Preliminaries} 
	\label{sec:preliminaries}
	\smallskip
	\noindent 
	\textbf{Graphs:} 
	In this contribution we consider directed as well as undirected
	connected graphs $G=(V,E)$ with vertex set $V(G)\coloneqq V$
	and edge set $E(G)\coloneqq E$.
	A graph is vertex or edge labeled if its vertices or
	edges are equipped with a labeling function respectively. If it is
	both vertex and edge labeled, we simply call the graph labeled. 
	We write $l(x)$ for the vertex labels $(x \in
	V(G))$ and edge labels $(x \in E(G))$.

	Given two (un)directed graphs $G$ and $G'$ and a bijection $\varphi:
	V(G) \rightarrow V(G')$, we say that $\varphi$ is edge-preserving if $(v,
	u) \in E(G)$ if and only if $(\varphi(v), \varphi(u)) \in E(G')$.
	Additionally, if $G$ and $G'$ are labeled, $\varphi$ is
	label-preserving if $l(v) = l(\varphi(v))$ for any $v \in V(G)$ and
	$l(v, u) = l(\varphi(v), \varphi(u))$ for any $(v, u) \in E(G)$. The
	bijection $\varphi$ is an isomorphism if it is edge-preserving
	and, in the case that $G$ and $G'$ are labeled, label-preserving. If
	$G = G'$, then $\varphi$ is also an automorphism.

	Given a (directed) graph $G$ we call $G$ (strongly) connected if
	there exists a path from any vertex $u$ to any vertex $v$. We call the
	subgraph $H$ of $G$ a (strongly) connected component if $H$ is a
	maximal (strongly) connected subgraph.

	Since the motivation of this work is rooted in chemistry,
	sometimes it
	is more natural to talk about the undirected labeled graphs as
	molecules, their vertices as atoms (with labels defining the atom
	type), and their edges as bonds (whose labels distinguish single,
	double, triple, and aromatic bonds, for instance), while still using
	common graph terminology for mathematical precision.

	\noindent
	\textbf{Graph Transformations:} 
	As molecules are modeled as undirected labeled graphs, it is
	natural to think of chemical reactions as graph transformations,
	where a set of educt graphs are transformed into a set of product
	graphs. 
	We model such transformations using the double pushout (DPO)
	approach.
	For a detailed overview of the DPO approach and its variations see
	\cite{habel2001double}. Here, we will use DPO as defined in
	\cite{andersen2016software} that specifically describes how to model
	chemical reactions as rules in a DPO framework.  

	A rule $p$ describing a transformation of a graph pattern $L$ into a
	graph pattern $R$ is denoted as a span $L \xleftarrow{l}{} K
	\xrightarrow{r}{} R$, where $K$ is the subgraph of $L$ remaining
	unchanged during rewriting and $l$ and $r$ are the subgraph
	morphism $K$ to $L$ and $R$ respectively. The rule $p$ can be
	applied to a graph $G$ if and only if (i) $L$ can be embedded in $G$
	(i.e., $L$ is subgraph monomorphic to $G$) and (ii) the graphs $D$ and
	$H$ exists such that the diagram depicted in \figname
	\ref{fig:derivation} commutes. 

	\begin{figure}[h]
		\centering
	\begin{tikzpicture}
		\node[] (L) at (0, 0){L};
		\node[] (K) at (1.5, 0){K};
		\node[] (R) at (3, 0){R};
		\node[] (G) at (0, -1.5){G};
		\node[] (D) at (1.5, -1.5){D};
		\node[] (H) at (3, -1.5){H};
		\draw[->] (K) -- node[above] {l} (L);
		\draw[->] (K) -- node[above] {r} (R);

		\draw[->] (L) -- node[right] {$m$} (G);
		\draw[->] (K) -- node[right] {} (D);
		\draw[->] (R) -- node[right] {} (H);

		\draw[->] (D) -- node[below] {l'} (G);
		\draw[->] (D) -- node[below] {r'} (H);
	\end{tikzpicture}
		\caption{A direct derivation.}
		\label{fig:derivation}
	\end{figure} 

	The graphs $D$ and $H$ are unique if they exist
	\cite{habel2001double}.  The graph $H$ is
	the resulting graph obtained by rewriting $G$ with respect to the rule
	$p$.  We call the application of $p$ on $G$ to obtain $H$ via the
	map $m: L \rightarrow G$, a
	direct derivation and denote it as $G \xRightarrow{p,m}{} H$ or $G
	\xRightarrow{p}{} H$, if $m$ is not important. We note, that $m$ is not necessarily 
	unique, i.e., there might exist a different map $m'$ such that $G \xRightarrow{p,m'}{} H$.

	For a DPO rule $p$ to model chemistry, we follow the modeling in \cite{chemicalmotifs}, and 
	impose 3 additional conditions that $p$ must 
	satisfy. (i) All graph morphisms must be injective (i.e., they describe subgraph isomorphisms).
	(ii) The restriction of graph morphisms $l$ and are $r$ to the vertices must be bijective, 
	ensuring atoms are conserved through a reaction.
	(iii) Changes in charges and edges (chemical bonds) must conserve the total number of 
	electrons.

\begin{figure} 
			\centering
			\includegraphics[width=1\textwidth]{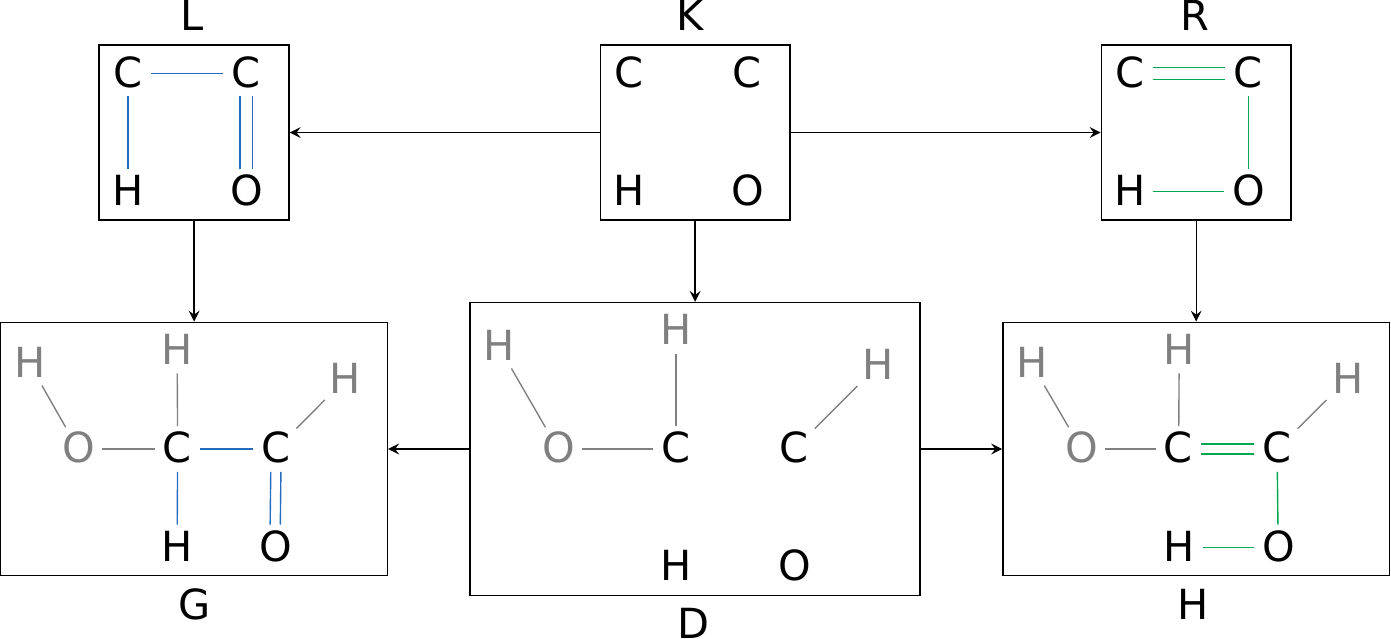}
			\caption{An example of a direct derivation. The mapping
			$l$, $r$, $l'$ and $r'$ is implicitly given by the
	depicted positions of the atoms. Given a chemical network, each
	hyper-edge directly corresponds to such a direct derivation.
			\label{fig:direct-der}
	}
	\end{figure}

	In the above framework, a chemical reaction is a direct derivation $G
	\xRightarrow{p,m}{} H$, where each connected component of $G$ and $H$
	corresponds to the educt and product molecules, respectively.
	Condition (i) and (ii), ensures that $l$ and
	$r$, and by extension $l'$ and $r'$ are bijective mappings when restricted to the vertices. As a
	consequence we can track each atom through a chemical reaction
	modeled as a direct derivation by the map $l'^{-1}\circ
	r'$. We note, that like $m$,  $l'$ and $r'$ might not be unique for a given direct derivation $G
	\xRightarrow{p}{} H$. We define the set of all such maps $l'^{-1}\circ
	r'$ for all possible maps $l'$ and $r'$ obtained from $G
	\xRightarrow{p}{} H$ as $tr(G \xRightarrow{p}{} H)$. An example
	of a direct derivation representing a chemical reaction is
	depicted in \figname \ref{fig:direct-der}.

	\noindent
	\textbf{Chemical Networks:}
	We consider a directed
	hypergraph where each edge $e=(e^+,e^-)$ 
	 is a pair of subsets of vertices. Moreover, we let $Y_e = e^+\cup e^-$
	denote the set of vertices that are comprised in the start-vertex $e^+$
	and end-vertex $e^-$ of $e$.  In short, a chemical network $\CN$ is a hypergraph where each vertex
	is a connected graph representing a molecule and
	each hyper-edge a rule application corresponding to a chemical
	reaction. Hence, every hyper-edge $e$ of $\CN$ corresponds to a set of
	direct derivations transforming the in-going vertices of $e$ into
	its out-going vertices. For a given set of edges $E$ of $\CN$, let
	$\mathcal{D}$ be the set of all direct derivations that can be
	obtained from $E$. Then, $tr(E) = \bigcup_{G\xRightarrow{p}{}
	H\in \mathcal{D}} tr(G\xRightarrow{p}{} H)$ and $tr(\CN) =
	tr(E(CN))$.\\

	\noindent
	\textbf{Semigroups and transformation semigroups:} 
	A \emph{semigroup} is a pair $(S, \circ)$, where $S$ is a set and $\circ: S
	\times S\rightarrow S$ an associative binary operator on $S$. 
	We often write $ab$ for the product $a\circ b$.  A semigroup that contains 
	the identity element $1$ (i.e.,  $s1 = s = 1s$ for all $s \in S$)
	is a \emph{monoid}. The \emph{order} of a semigroup $S$ is its cardinality $|S|$. A
	subset $A \subseteq S$ is said to \emph{generate $S$} or called a
	generating set for $S$, $\langle A \rangle =
	S$, if all elements of $S$ can be expressed as a finite product of
	elements in $A$.

	Given a non-empty finite set $X$, a \emph{transformation} on $X$
	is an arbitrary map $f: X \rightarrow X$ that assigns to \emph{every} element
	$x\in X$ some element $f(x)\in X$. The identity of a
	transformation on $X$ is denoted $1_X$. A transformation monoid
	is a transformation semigroup
	with identity. If $X=\{1,\dots,n\}$, we
	often use the notation $(i_1, i_2, \dots, i_n)$ for the
	transformation $f(j)=i_j$, $1\leq j\leq n$. Note, the elements
	$i_1, i_2, \dots, i_n$ need not necessarily be pairwise distinct.
	Let $T$ be the set of all possible transformations on $X$. If $S\subseteq T$
	and $S$ is closed under function composition $\circ$, then $(S,\circ)$ forms a
	semigroup, also called a \emph{transformation semigroup}. To
	emphasize that $S$ is a collection of transformations on $X$, we
	will use the notation $(X,S)$ for transformation semigroups and
	say that $S$ \emph{acts on} $X$.
	Given a tuple $\bar{z} = (z_1, z_2, \dots, z_n)$ of $n$ distinct
	elements of $X$ and a transformation semigroup  $(X,S)$, the
	orbit of $\bar{z}$ is defined as $\mathcal{O}(\bar{z}, S) =
	\{(s(z_1), \dots, s(z_n)\mid s \in S\}.$ In what follows, we
	use the notion $y\in t= (i_1, i_2, \dots, i_n)$ to indicate that
	$y = i_j$ for some $j$, $1\leq j\leq n$.  

	Given a transformation semigroup $(X,S)$ with generating set $A$, in
	symbols $S=\langle A \rangle$, we will employ the \emph{(right) Cayley
	  graph} $\cay(S, A)$ of $S$ and $A$ with vertex set $S$ and edge set
	$E(\cay(S, A)) = \{(s, sa) \mid s\in S, a\in A\}$.  In addition, every
	edge $(s, sa)$ of $\cay(S, A)$ obtains label $l_a$, that is, the
	unique label that is associated to each generator $a$ in $A$.
	Similarly, the \emph{projected} Cayley graph $\pcay(S, A, \bar{z})$ is
	defined for tuples $\bar{z}$: It has vertex set
	$\mathcal{O}(\bar{z}, S)$ and for all $s\in \mathcal{O}(\bar{z}, S)$
	and for all $a\in A$ there is an edge $(s, sa)$ with label $l_a$.
	A \emph{free semigroup} $\Sigma^+$, is the semigroup containing all
	finite sequences of strings constructed from the alphabet $\Sigma$
	with concatanation as the associative binary operator. Adding the
	empty string $\epsilon$ results in the free monoid $\Sigma^* = \Sigma^+ \cup \{\epsilon\}$.

	\section[Chemical Networks and Algebraic Structures]{Chemical Networks and their Algebraic Structures}
	\label{sec:chemical-networks-and-algebraic-structures}
	\subsection{Characteristic Monoids}\label{sec:charsg}
	Assume we are given some chemical network $\CN$ that is some hypergraph
	modeling some chemistry. 
	As we are interested in tracking the possible movements of atoms
	in $\CN$, we are inherently interested in the reactions of $\CN$,
	i.e., in its edge set $E(\CN)$. Indeed, atoms can only reconfigure to construct new
	molecules under the execution of some reaction. We will refer to
	the execution of a reaction as an \emph{event}. The possible
	reconfigurations of atoms caused by a single event, is given by
	the set of atom maps $tr(\CN)$ constituting a set of (partial)
	transformations on $X = \bigcup_{M \in V(\CN)}V(M)$. Note, the
	vertex $M\in V(\CN)$ corresponds to an entire molecule for which
	$V(M)$ denotes the set of atoms (=labeled vertices). 
	A transformation $t$ on $X$ describes the position 
	(i.e., in what molecule and where in the molecule the atom is found)
	 of each atom in $X$ when $X$ is transformed by $t$.
	 In what follows, we will sometimes refer to such
	transformations on $X$ as \emph{atom states}, as the
	transformations encapsulates the "state" of the network, i.e., the position of each atom.
	To track the possible movement of atoms through a
	chemical network, we must consider sequences of events.

	\begin{definition}[Event Traces]
			Let $\Sigma$ be an alphabet containing a unique
			identifier $t$ for each atom map in $tr(\CN)$. Then, an
			\emph{event trace} is an element of the free
					monoid
			$\Sigma^*$.  
	\end{definition}

	The free monoid $\Sigma^*$ contains all possible sequences of
	events that can move the atoms of $X$. Note, $\Sigma^*$ does not
	track the actual atoms through event traces. For this, we use the
	following structure:

	\begin{definition}[Characteristic Monoids] \label{def:charsg} Let the
	  characteristic monoid of $\CN$ be defined as the transformation
	  monoid $ \charsg = (X, \langle tr(\CN) \cup 1_X \rangle).$ Moreover,
	  given a set of edges $E \subseteq E(\CN)$, and the set of atoms
	  $Y \subseteq X$ found in $E$ (that is $Y= \cup_{e\in E} Y_e$), we
	  let the characteristic monoid of $E$ be defined as
	  $ \mathcal{S}(E) = (Y, \langle tr(E) \cup 1_Y \rangle).  $

	\end{definition}

	Let $\sigma: \Sigma \rightarrow tr(\CN)$ be the function, that
	maps all identifiers of $\Sigma$ to their corresponding atom map
	in $tr(\CN)$. Given an event trace $t = t_1t_2\dots t_n \in
	\Sigma^*$, we let the events of $t$ refer to their corresponding
	transformations in $tr(\CN)$ when acting on an element $s\in\charsg$, i.e., $st =
	s\sigma(t_1)\sigma(t_2)\dots \sigma(t_n) \in \mathcal{S}(\CN)$.
	Every event trace $t \in \Sigma^*$ gives rise to a
	member $\charsg$, in particular the transformation $1_Xt$, that represents the
	resulting atom state obtained from moving atoms according to $t$. Hence,
	there is a homomorphism from $\Sigma^*$ to $\charsg$, meaning
	that $\charsg$ captures all possible movements of atoms through
	reactions of $\CN$.

	Often, we are only interested in tracking the movement of a small
	number of atoms. Let $\bar{z}$ be a tuple of distinct elements from
	$X$ that we want to
	track. 
	Then, there is again a homomorphism from $\Sigma^*$
	and $\orbit(\bar{z}, \charsg)$. Namely, for a given event trace
	$t \in
	\Sigma^*$, we can track the atoms of $\bar{z}$ as the atom state
	$1_{\{x\ |\ x \in \bar{z}\}}t$ corresponding to an
	element in the orbit $\orbit(\bar{z}, \charsg)$, if we treat the
	element as a (partial) transformation.
	As a result, $\orbit(\bar{z}, \charsg)$
	characterizes the possible movements of the atoms in $\bar{z}$,
	and we will refer to its elements as atom states similarly to
	elements in $\charsg$ as they conceptually represent the same
	thing.

	We note, the above definitions are not unlike some of the core
	definitions within algebraic automata theory
	\cite{algebraic-automata}. Here, the possible inputs of an
	automata is often defined in terms of strings obtained from the
	free monoid on the alphabet of the automata. The characteristic
	semigroup is then defined as the semigroup that
	characterizes the possible state transitions. In
	the same vein, we
	can view our notion of event traces as the possible "inputs" to
	our chemical network CN that moves some initial configuration of
	atoms $1_X$. The characteristic monoid of CN
	then characterize the possible movements of atoms through event
	traces.

	In what follows we let $\cay(\CN)$ denote the Cayley graph
	$\cay(\charsg$, $tr(\CN) \cup 1_X)$. 
	Similarly, given a tuple of atoms
	$\bar{z}$, we let $\pcay(\CN, \bar{z})$ denote the projected
	Cayley graph $\pcay(\charsg, tr(\CN)\cup 1_X, \bar{z})$. We note,
	that by \defname \ref{def:charsg}, $\charsg$ is constructed
	from the generating set $\langle tr(\CN)\cup 1_X \rangle$, and
	hence $\cay(\CN)$ and $\pcay(\CN, \bar{z})$ are well defined.
	Since
	the transformation $1_X$ will always result in a loop on every
	vertex of the (projected) Cayley graph, and conveys no meaningful
	information, we will refrain from including any edge arising from
	$1_X$.

	We can illustrate the relation between atom states using the Cayley
	graph $\cay(\CN)$. More precisely, there exists an edge between
	two atom states $a,b \in \charsg$ with label $t$, if it is
	possible to move the atoms in $a$ to $b$ using $t$. It is natural
	to relate $\Sigma^*$ to $\cay(\CN)$. Namely, any path in
	$\cay(\CN)$ corresponds directly to an event trace in
	$\Sigma^*$.  
	Hence, where $\Sigma^*$ encapsulates the "inputs" of the chemical
	network and $\charsg$ contains the possible atom states derived from
	$\Sigma^*$, the Cayley graph $\cay(\CN)$ captures
	\emph{how} atom states from $\charsg$ can be created by event
	traces.
	\begin{figure}[t]
		\centering
			\centering
	\subfloat[] {
			\includegraphics[width=0.29\textwidth]{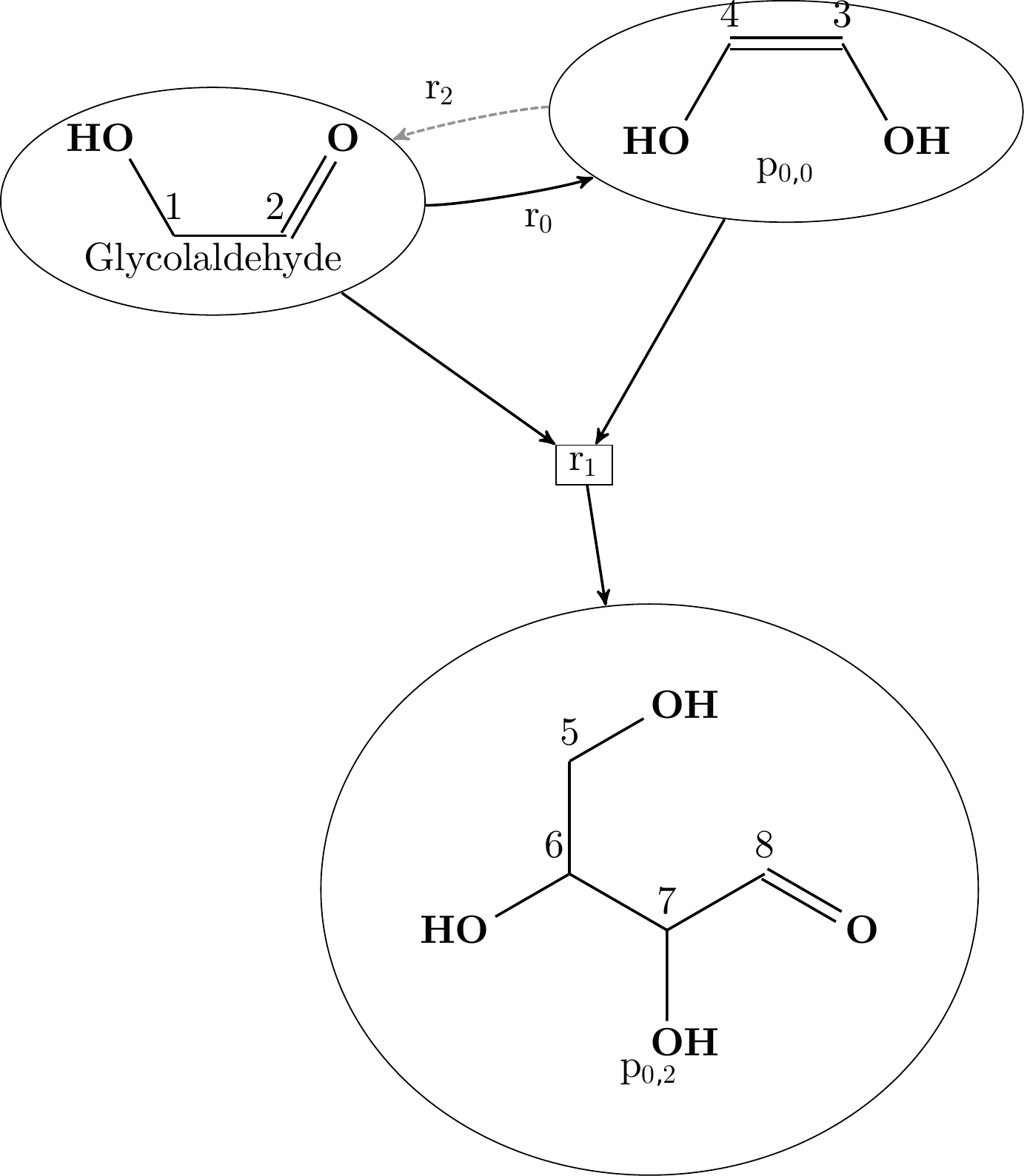}
			\label{fig:formose-example}
	}
	\subfloat[] {
			\raisebox{1.5cm}{
			\includegraphics[width=0.6\textwidth]{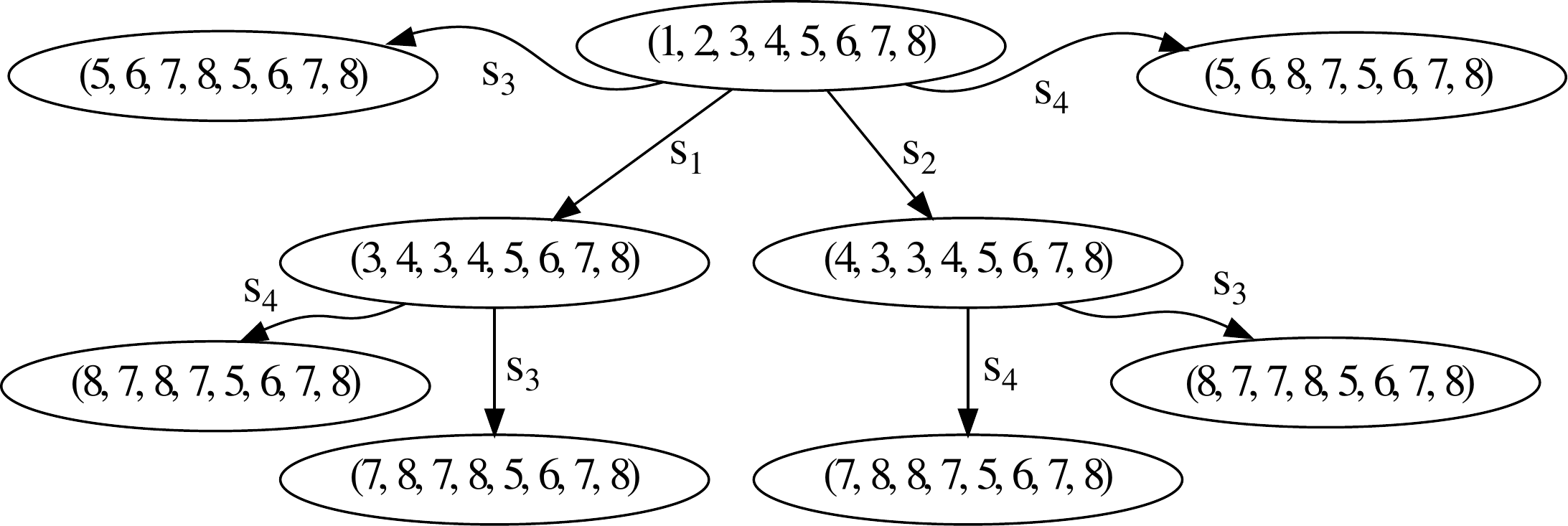}
			\label{fig:formose-Cayley}
	}
	}
		\caption[]{\subref{fig:formose-example} A small example using
				molecules and reactions found in the Formose
				reaction. The carbon atoms of each molecule are
				labeled with a unique identifier for easy reference.
			\subref{fig:formose-Cayley} The Cayley graph $\cay(\CN)$
			of \figname \ref{fig:formose-example} from the example of
			Sec.~\ref{sec:charsg}.
			From the graph, we see the longest path from $1_X$ has length
			$2$, meaning that any event trace can at most transform
			$1_X$ meaningfully twice. In fact, only two types of
			event traces are of interest: Either the tracked atoms
			are immediately moved by the reaction $r_1$ to $p_{0,2}$,
	or the atoms of glycolaldehyde are first moved to $p_{0,0}$ using
	$r_0$, and then moved to $p_{0,2}$.
	}
	\end{figure}

	\smallskip
	\textbf{Example:} \label{sec:ex_simple}
	As an illustrative example, consider the reaction network $\CN$
	depicted in \figname\ref{fig:formose-example}. For simplicity we will use
	 reactions $r_0$ and $r_1$ involved in the so-called Formose reaction. We
	 restrict ourselves to
	only consider the carbon atoms of all molecules, and have labeled them
	with a corresponding unique id for easy reference. Here, the underlying set 
	$X=\{1,2,\dots, 8\}$ corresponds to the eight elements labeled by
	$1,2,\dots, 8$ in  \figname \ref{fig:formose-example}.
	 From $tr(\CN)$ we get 4 
	transformations: 
		$s_1 = [3,4,3,4,5,6,7,8]$, 
		$s_2 = [4,3,3,4,5,6,7,8]$ (both obtained from $r_0$), and
		$s_3 = [5,6,7,8,5,6,7,8]$, 
		$s_4 = [5,6,8,7,5,6,7,8]$ (both obtained from $r_1$)
	with the corresponding alphabet $\Sigma = \{s_1, s_2, s_3, s_4
	\}$.
	For a reaction, the corresponding transformation(s) maps the atoms of
	the educt molecules to the atoms of the product molecules while all
	other atoms are mapped with the identity.  The transformations
	describe how carbon atoms are rearranged into different configurations
	when an event is fired. $s_1$ and $s_2$ describe how the carbon atoms
	of a glycolaldehyde molecule are arranged in the molecule $p_{0,0}$
	when transformed via the reaction $r_0$. In the case of $s_1$, we see
	that the carbons are rearranged such that $s_1(1) = 3$ and
	$s_1(2) = 4$. Of course, due to the symmetries in the molecule
	$p_{0,0}$, reaction $r_0$ also results in the mirrored transformation
	of $s_1$, i.e., $s_2(1) = 4$ and $s_2(2) = 3$.  The characteristic
	monoid of $\CN$, $\mathcal{S}(\CN)$, has an order of $9$. We
	illustrate the movement of atoms in $\CN$ by its Cayley graph
	$\cay(\CN)$ which is depicted in \figname \ref{fig:formose-Cayley}.  Any
	path originating from the identity element corresponds to an
	event trace, e.g. we can track the atoms $1$ and $2$ through the event
	trace $s_1s_3$ as the corresponding path and realize $s_1s_3(1) = 8$
	and $s_1s_3(2) = 7$.
	\begin{figure}[t]
			\centering
	\subfloat[] {
		\centering
		\includegraphics[width=0.24\textwidth]{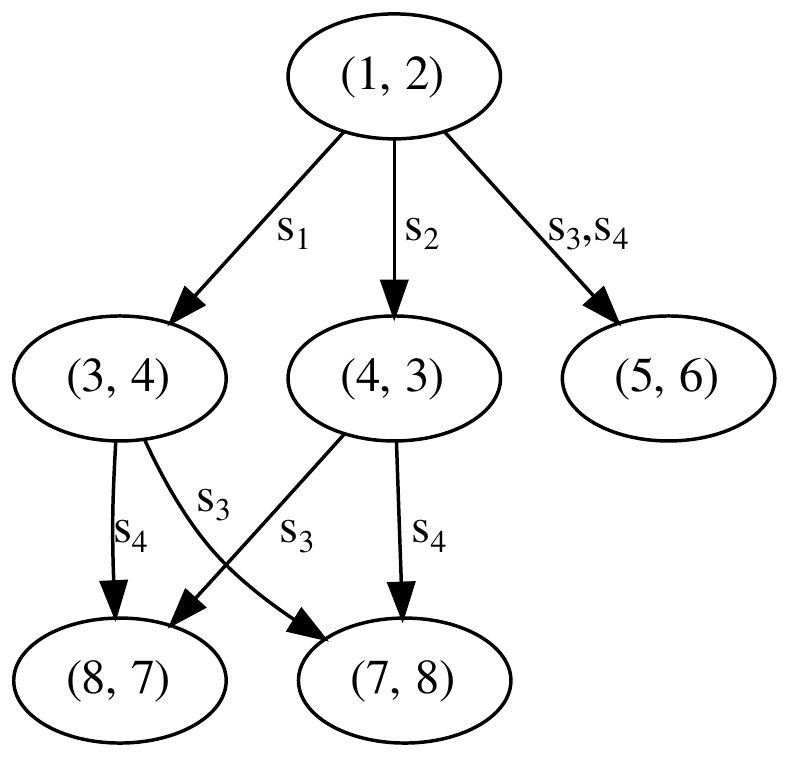}
		\label{fig:formose_orbit_graph}
	}
	\subfloat[] {
			\raisebox{0.5cm}{
		\includegraphics[width=0.45\textwidth]{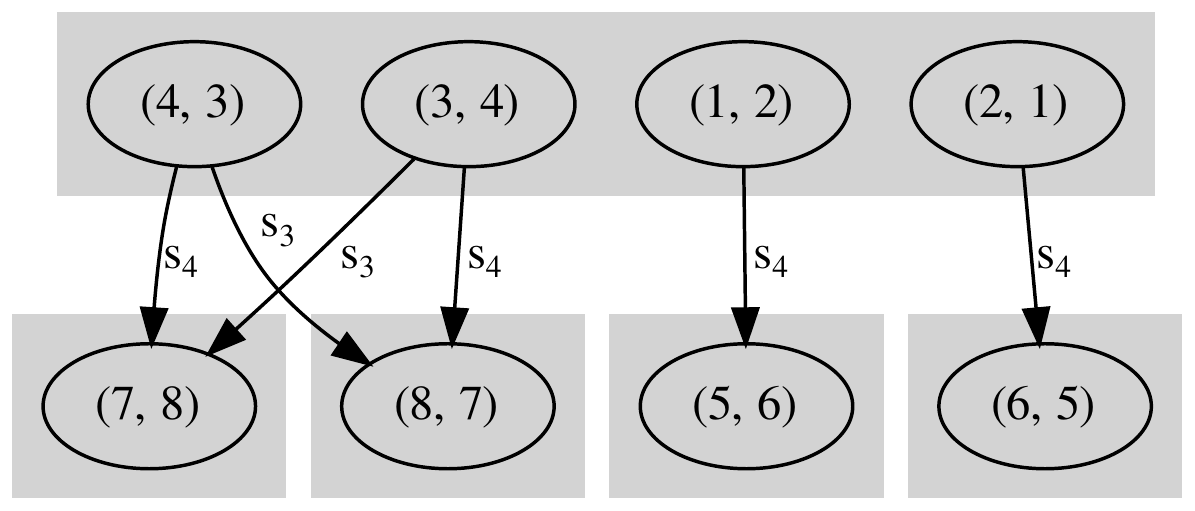}
			\label{fig:example-cycle}
		}
		}
	\caption[]{
	 \subref{fig:formose_orbit_graph} The projected Cayley graph
	 $\pcay(\CN, (1,2))$ from the example of Section
	 \ref{sec:charsg}.
	 Like for $\cay(\CN)$, we see there are only two types
	of event traces of interest. However, since we are only tracking
	the atoms of the glycolaldehyde molecule, some atom states are
	effectively coalesced compared to $\cay(\CN)$.
				\subref{fig:example-cycle}
				The projected Cayley graph $\pcay(\CN, (1,2))$ from
				the example of Sec.~\ref{sec:subsystem}.
				The graph shows the natural subsystems of the
							carbon atoms of a glycolaldehyde molecule.
							Vertices in the same box constitutes vertices
							that are in the same natural subsystem. Note,
							edges between vertices in the same natural
							subsystem are not depicted (e.g, one of the
							eight hidden edges in the top-level subsystem
							is $(3,4)\rightarrow (1,2)$ with label $s_5$).
						  }
	\end{figure}
	Assume now, that we were only interested in tracking the carbon atoms
	found in the glycolaldehyde molecule. To this end, we can examine
	$\orbit(\bar{z}, \charsg)$, which contains $6$ elements, meaning
	there exists $6$ unique atom states for the atoms in a
	glycolaldehyde molecule.  Again, we
	can study these movements using the projected Cayley graph
	$\pcay(\CN, (1,2))$. The resulting graph is
	depicted in \figname \ref{fig:formose_orbit_graph}.

	\subsection{Natural Subsystems of Atom States} \label{sec:subsystem}
	In the intersection between group theory and systems
	biology, attempts to formalize the notion of
	natural subsystems and hierarchical relations within such systems
	have been done by works such as \cite{nehaniv2015symmetry}.
	Here, natural subsystems are defined as symmetric structures
	arising from a biological system. Such symmetries
	manifests as permutation groups of the associated semigroup
	representing said system. In such a model the Krohn-Rhodes
	decomposition or the holonomy decomposition
	\cite{egri2015computational} can be used to construct a hierarchical
	structure on such natural subsystems of the biological system.
	In terms of atom tracking, however, defining natural subsystems
	in terms of the permutation groups in $\charsg$ does not have an
	immediately useful interpretation. Similarly, the
	hierarchical
	structure obtained from methods such as holonomy decomposition are not intuitive to interpret.
	Instead, when talking about natural subsystems in terms of atom
	tracking, we are interested in systems of reversible
	\emph{event traces}, i.e., event traces that do not change the
	original configuration of atoms.
	To this end, it is natural to define natural subsystems of
	$\charsg$ in terms of Green's relations \cite{algebraic-theory-semigroup}.
	For elements $s_1, s_2 \in \charsg$, we define the reflexive
	transitive  relation $\succeq_{\mathcal R}$ as $s_1 \succeq_{\mathcal R} s_2$, if
	there exists an event trace $t \in \Sigma^*$ such that $s_1 t = s_2$.
	In addition, we define an equivalence relation $\mathcal{R}$, 
	where  $s_1$ is equivalent to $s_2$, in symbols  $s_1\mathcal{R}s_2$  
	whenever $s_1 \succeq_{\mathcal{R}} s_2$ and $s_2 \succeq_{\mathcal{R}} s_1$. 
	\begin{definition}[Natural Subsystems]
			The natural subsystems of $\charsg$ is the set of
			equivalence classes induced by the
			$\mathcal{R}$-relation.
	\end{definition}
	The equivalence classes correspond to the strongly connected components of the Cayley
	graph $\cay(\CN)$ \cite{froidure1997algorithms}.
	We note,
	that for a tuple of atoms $\bar{z}$,
	 the natural extension to natural subsystems of the orbit
	 $\orbit(\bar{z}, \charsg)$ is simply the strongly connected
	 components of its projected Cayley graph $\pcay(\CN,
	 \bar{z})$.
	The $\mathcal{R}$ relation is interesting, as the equivalence classes
	on $\charsg$ induced by the $\mathcal{R}$ relation forms pools
	of reversible event traces. 
	More precisely, let $s_1\mathcal{R} s_2$ for some $s_1, s_2 \in
	\charsg$, where $s_1\cdot t_{12} = s_2$ and $s_2\cdot t_{21} =
	s_1$ for some $t_{12},t_{21}\in \Sigma^*$.
	Then, the event traces $t_{12}$ and $t_{21}$ are reversible, i.e.
	we can re-obtain $s_1$ as $s_1t_{12}t_{21} = s_1$ and $s_2$ as
	$s_2t_{21}t_{12} = s_2$.
	Additionally, the quotient graph of the
	equivalence classes of the $\mathcal{R}$ relation on the Cayley
	graph $\cay(\CN)$ naturally forms a hierarchical relation on the
	atom states of $\charsg$ that has a useful
	interpretation from the point of view of chemistry as we will see
	in Sec.~\ref{SEC:TCA}.

	\textbf{Example: }
	Again, consider the reaction network obtained from the formose
	reaction depicted in \figname \ref{fig:formose-example}. We will include
	the transformations obtained from reaction $r_2$ in additions to the
	transformations listed in Sec.~\ref{sec:ex_simple}:
	$s_5 = [1,2,1,2,5,6,7,8]$ and
	$s_6 = [1,2,2,1,5,6,7,8]$ (both obtained from $r_2$).
	Assume we are interested in determining how carbon atoms of a
	glycolaldehyde molecule can reconfigure into different molecules.
	The projected Cayley graph $\pcay(\CN, (1,2))$ shows 
	such configurations and is depicted in
	\figname \ref{fig:example-cycle}. Here, the atom states belonging to the
	same gray box are strongly connected and hence belong to the same
	natural subsystem. For clarity, we
	have removed edges between atom states in the same subsystem, since
	any atom state in a subsystem can be transformed into any other state
	in the same subsystem.

	Notably, we see from \figname \ref{fig:example-cycle} that the atoms $1$
	and $2$ in the glycolaldehyde molecule can swap positions. We could of
	course also realize that such a swap was possible by noticing the
	symmetries in the glycolaldehyde molecule and the fact that we can
	convert glycolaldehyde to the $p_{0,0}$ molecule and vice versa.
	However, such patterns becomes immediately obvious from the
	projected
	Cayley graph.
	Finally, we can derive from \figname \ref{fig:example-cycle}, that
	it is only possible to leave the original subsystem by applying
	transformation $s_3$ or $s_4$, corresponding to reaction~$r_1$. 

	\section{Results}
	\subsection{Implementation}
	To test the practicality of the structures introduced in the 
	previous section, we implemented the construction of the 
	projected Cayley graph of a set of atoms in a chemical network.
	The resulting implementation can be found at
	\url{https://github.com/Nojgaard/cat}
	All code is written in python and uses the software package MØD 
	\cite{andersen2016software} and NetworkX \cite{SciPyProceedings_11}
	to construct the chemical networks 
	and find the transformations used for the characteristic monoid.
	All figures in the following section were constructed with said 
	implementation, and each run finished within seconds on an
	8 core Intel Core i9 CPU with 64 GB memory. The most time consuming part of the implementation
	was the computation of the transformations obtained from each hyper-edge in
	the chemical network. In contrast, the construction time of the projected Cayley graph
	proved to be negligible.

	\label{sec:results}
	\subsection{Differentiating Pathways}
	In this section, we will explore the possibilities of using the
	characteristic monoids of chemical networks to determine
	if it is possible to distinguish between two pathways $P_1$ and
	$P_2$, based on their atom states of their respective
	characteristic monoids.
	The motivation stems from methods
	such as isotope
	labeling.
	Here, a "labeled" atom, is a detectable isotope whose position is
	known in some initial molecule and can then be detected, along with
	its exact position, in the product molecules of some pathway.
	In contrast to \cite{IsotopeLabel2019}, we will not focus on the
	orbits of atoms in isolation, as we lose the ability to reason about
	atom positions in relation to each other.
	Moreover, as we will see here, the Cayley
	graph of the chemical network can be used to identify the exact event
	two pathways split.

	Given a chemical network $\CN$, a pathway $P$ is a set
	of hyper-edges (i.e. reactions)
	from $\CN$ equipped with a set of input and output molecules. 
	We think of a pathway
	as a process that consumes a set of input molecules to
	construct
	a set of output molecules, using the reactions specified by
	$P$.
	In our case, a "labeled" atom is a point in $\charsg$. 
	Given two pathways $P_1$ and $P_2$, we can characterize the
	possible movement of atoms as the characteristic monoids
	$\mathcal{S}(P_1)$ and $\mathcal{S}(P_2)$.
	In practice, it might not be
	feasible to track every atom in $\CN$, e.g. we are only able to
	replace a few atoms with its corresponding detectable isotope,
	and hence it becomes useful to consider the orbits
	$\orbit(\bar{z}, \mathcal{S}(P_1))$ and $\orbit(\bar{z},
	\mathcal{S}(P_2))$ where
	$\bar{z}$ is the atoms from the input molecules we can track.
	Clearly, of the atom states in $\orbit(\bar{z}, \mathcal{S}(P_1))$ and $\orbit(\bar{z},
	\mathcal{S}(P_2))$, we can only expect to observe, e.g. in
	an isotope labeling experiment, the atom
	states that locates the tracked atoms in the output molecules.
	As a result, we arrive at the following observation:

	\begin{observation}
	  Let $Y_i \subseteq \orbit(\bar{z}, \mathcal{S}(P_i))$,
	  $i \in \{1,2\}$, be the atom states we can hope to observe after
	  some isotope labeling experiment. Then, we can always distinguish
	  between $P_1$ and $P_2$ if $Y_1\cap Y_2 = \emptyset$.
	\end{observation}

\begin{figure}[t]
			\centering
			\subfloat[] {
			\includegraphics[width=0.75\textwidth]{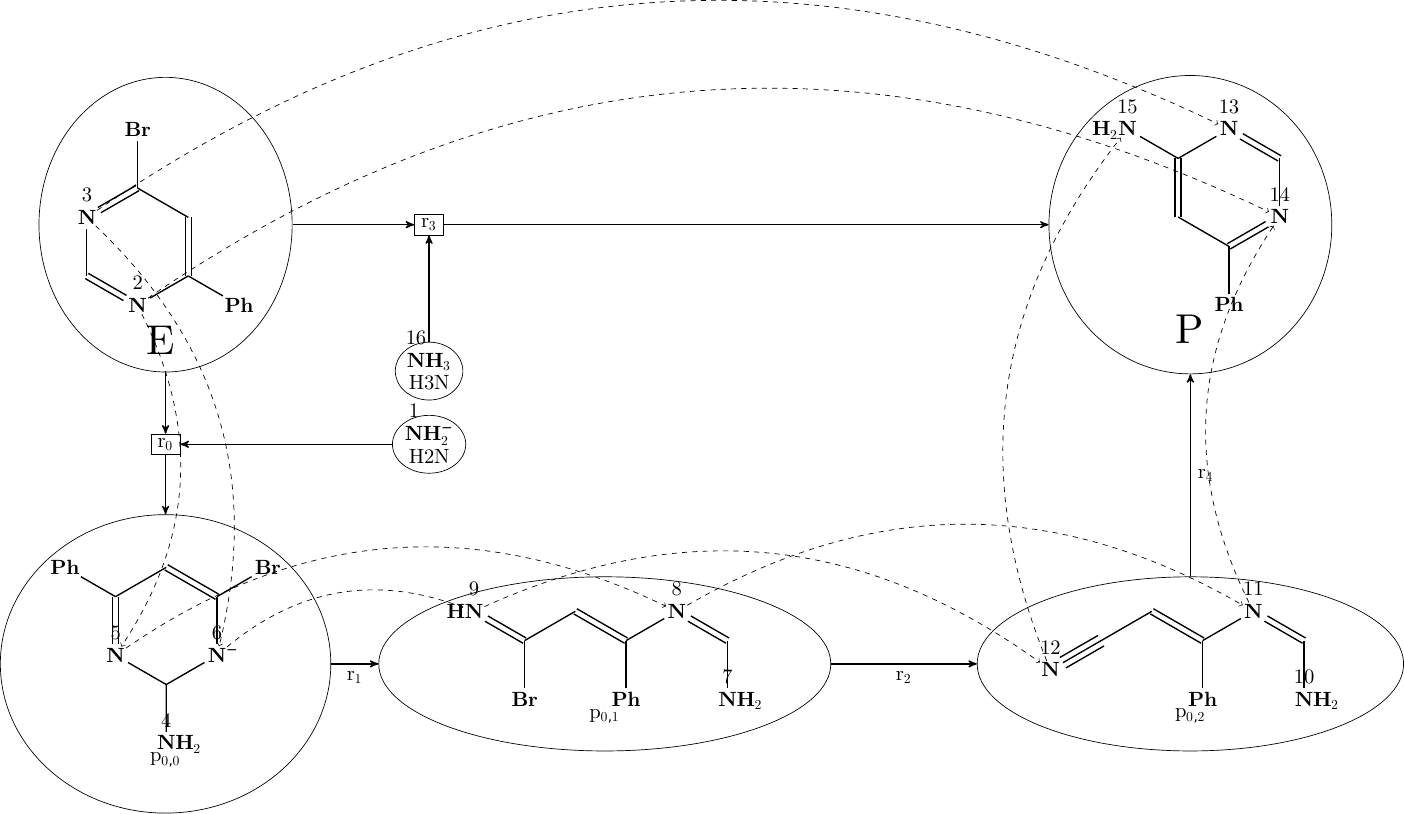}
			\label{fig:anrorc-dg}
			}
			\subfloat[] {
			\includegraphics[width=0.25\textwidth]{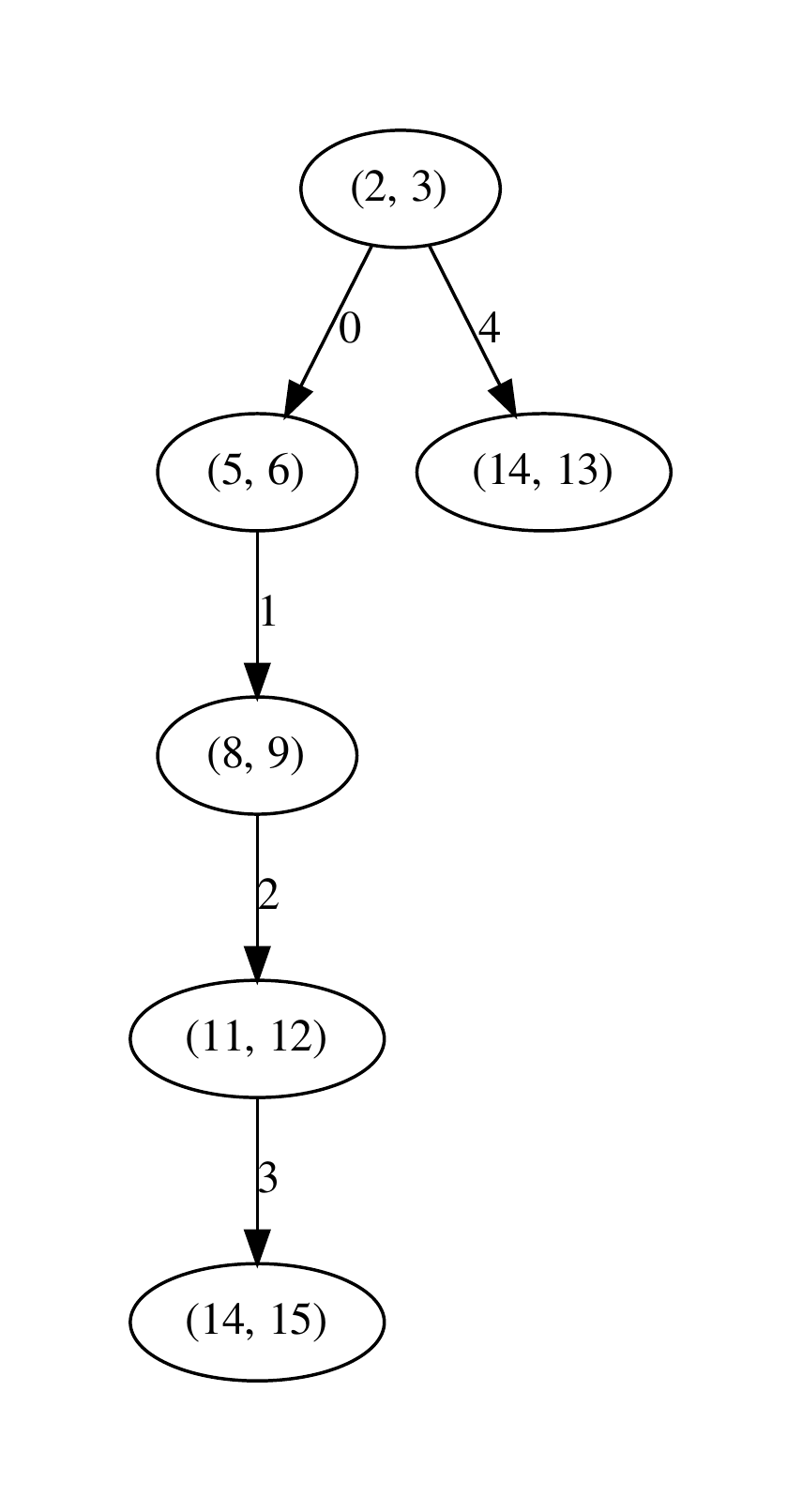}
			\label{fig:anrorc-cay}
			}
			\caption[]{\subref{fig:anrorc-dg} The chemical network
					  for the creation of P from E using
					  ammonia. The dotted lines shows the
					  possible atom trajectories for the atoms $2$ and $3$
					  respectively. \subref{fig:anrorc-cay} The projected
					  Cayley graph $\pcay(\CN, (2,3))$.
	}
	\end{figure}

	\noindent
	\textbf{Example:} Consider the network $\CN$
	depicted in \figname \ref{fig:anrorc-dg} modelling the creation of
	product 4-phenyl-6-aminopyrimidine (denoted P) from the educt
	4-(benzyloxy)-6-bromopyrimidine (denoted E) using ammonia. This well
	investigated and widely used substitution mechanism (ANRORC) \cite{anrorc} was
	proven to non-trivially function via ring opening and ring closure
	(and an accompanied carbon replacement) via isotope labeling.
	Two possible pathways are modelled: the input molecules for the two
	pathways are the molecules E, \ch{NH3}, \ch{NH2}, while the output is the
	single molecule P. The first, seemingly correct but wrong, pathway
	$P_1 = \{r_3\}$ converts E and an \ch{NH3} molecule directly into P, by
	replacing the \ch{Br} atom with \ch{NH2}. The second pathway consists of the
	reactions $P_2 = \{ r_0, r_1, r_2, r_4 \}$ and models the ANRORC
	mechanism.

	Assume we wanted to
	device a strategy to decide what pathway is executed in
	reality. By replacing the nitrogen atoms of the E molecule with
	the isotope $^{13}$N we would be able to observe where the atoms are
	positioned in the produced P molecule. 
	Since we, by assumption, only label the nitrogen atoms of the
	E molecule, i.e., the atoms $3$ and $2$, we can look at the
	orbits of the
	characteristic monoids $\orbit((2,3), \mathcal{S}(P_1))$ and
	$\orbit((2,3), \mathcal{S}(P_2))$ with the order of $5$ and $2$
	respectively. We see that both orbits only contains a single
	element locating $(2,3)$ in the $P$ molecule, namely the
	element $(14,15)$ for $\orbit((2,3), \mathcal{S}(P_1))$ and $(14,13)$
	for $\orbit((2,3), \mathcal{S}(P_2))$. As the possible configurations
	are different for $P_1$ and $P_2$, it is hence possible to always
	identify if the $P$ molecule was created by $P_1$ or $P_2$.

	This fact, also becomes immediately obvious by looking at the
	projected
	Cayley graph $\pcay(\CN, (2,3))$ depicted in \figname
	\ref{fig:anrorc-cay}, that shows the immediate divergence of atom
	states of the two pathways.

	\subsection{Natural Subsystems in the TCA Cycle}\label{SEC:TCA}
	The citric acid cycle, also known as the tricarboxylic
	(TCA) cycle or the Krebs cycle, is at the heart of many metabolic
	systems. The cycle is used by aerobic organisms to release stored
	energy in the form of ATP by the oxidation of acetyl-CoA into
	water and CO$_2$. The details for the TCA cycle can be found in
	any standard chemistry text book, e.g. \cite{biochem-textbook}.
	In \cite{smith2016origin}, the
	trajectories of different carbon atoms in the TCA cycle was examined to explain the
	change of their oxidation states. 
	It is well known that there is an enzymatic differentiation
	of the two carboxymethyl groups in citrate, which requires a rigorous
	stereochemical modeling of the graph grammar rules used
	\cite{stereoGraTra}. Ignoring such stereochemical modeling would lead
	to atom mappings not occurring in nature. We will provide a formal handle to analyze
	theoretically possible carbon trajectories using the algebraic constructs
	provided in this paper. As we will see, such structures provides
	intuitive interpretations for the TCA cycle. 
	More precisely, assume we are
	interested in answering the following questions: What are the possible
	trajectories of the carbons of an oxaloacetate (OAA) molecule within the TCA cycle
	while i.) ignoring the enzymatic differentiation
	of the two carboxymethyl groups in citrate (denoted TCA-$\square$), or ii.) not ignoring (denoted TCA-${\tetrahedron}$).
	 To answer these questions,
	we will decompose the characteristic monoid of the TCA cycle into its
	natural subsystems and examine them using the projected Cayley graph.

\begin{figure}
		\centering
		  \includegraphics[width=0.8\textwidth]{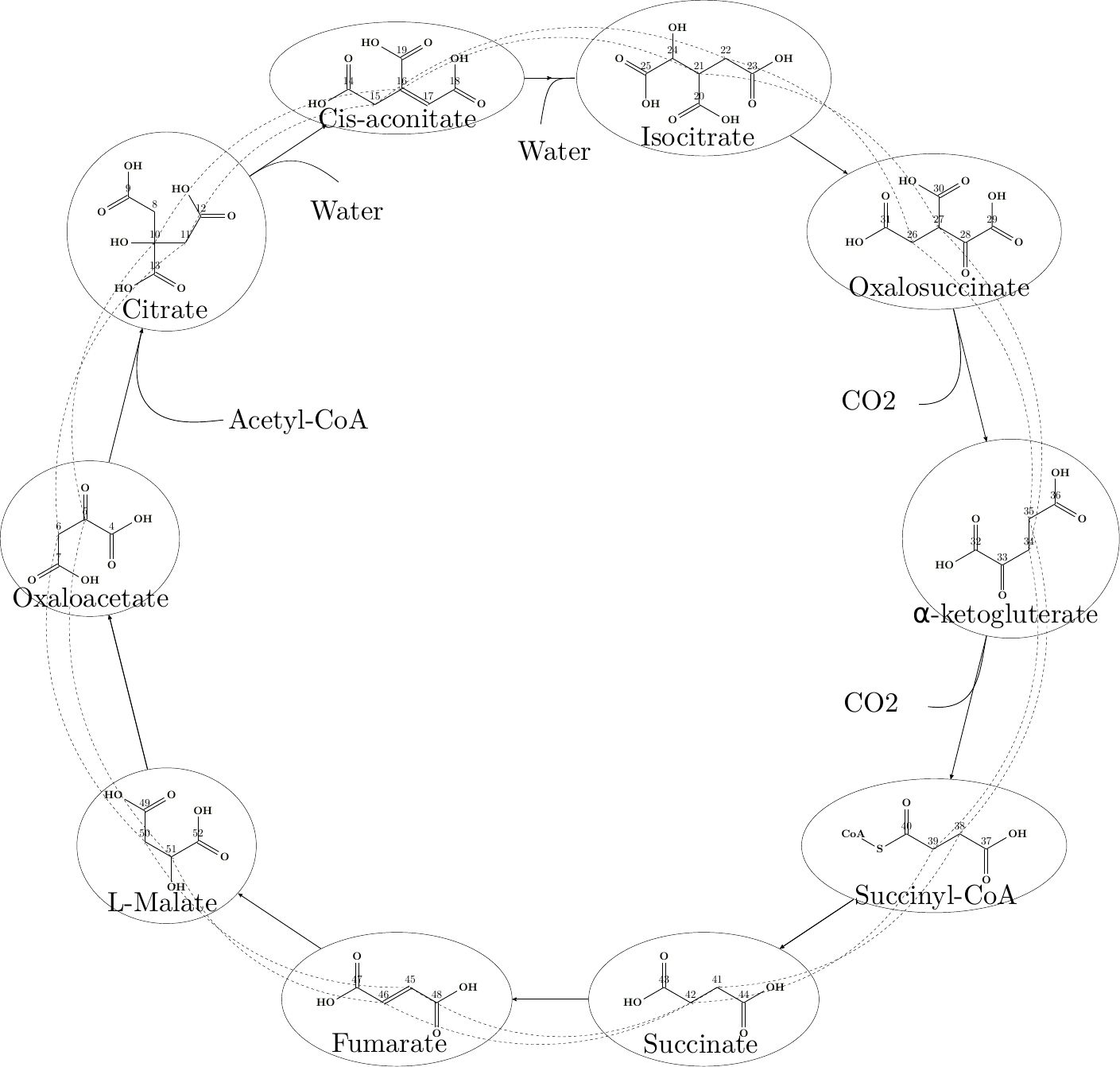}
		  \caption[]{
		A (simplified) chemical network modelling the TCA cycle.
		Note, any molecules not containing carbon atoms are modelled,
		but not depicted here. Each carbon atom is equipped with a
		unique id for easy reference.   
		  }
		  \label{fig:krebbs_cay_path}
	\end{figure}

	In our setting, the TCA cycle is the chemical network $\CN$,depicted in
	\figname \ref{fig:krebbs_cay_path},  giving rise
	to transformations of the underlying monoid.
	The network is
	made up of 13 reactions, however, some of the reactions are not
	shown for
	simplicity. Of these 13 reactions, 7 of them yields exactly 1
	transformation each while the remaining 6 yields 2 possible
	transformations each, resulting in a total of 19 transformations
	found. The reactions containing multiple transformations are due
	to automorphisms in molecules
	such as citrate and fumarate. When the enzymatic differentiation of
	the carboxymethyl group in citrate is not ignored, only 4 of the
	13 reactions yield 2 possible transformations, as the carbon
	traces to and from citrate are more constrained. In short, while
	both $\TCA$ and $\sTCA$ are modeled by the same network, the
	obtained transformations differ. More precisely, $|tr(\CN)| =
	19$ wrt. $\TCA$ and $|tr(\CN)| = 17$ wrt. $\sTCA$.

\begin{figure}[t]
		\centering
		\subfloat [] {
			\label{fig:krebbs_cay_full}
			\includegraphics[width=0.5\textwidth]{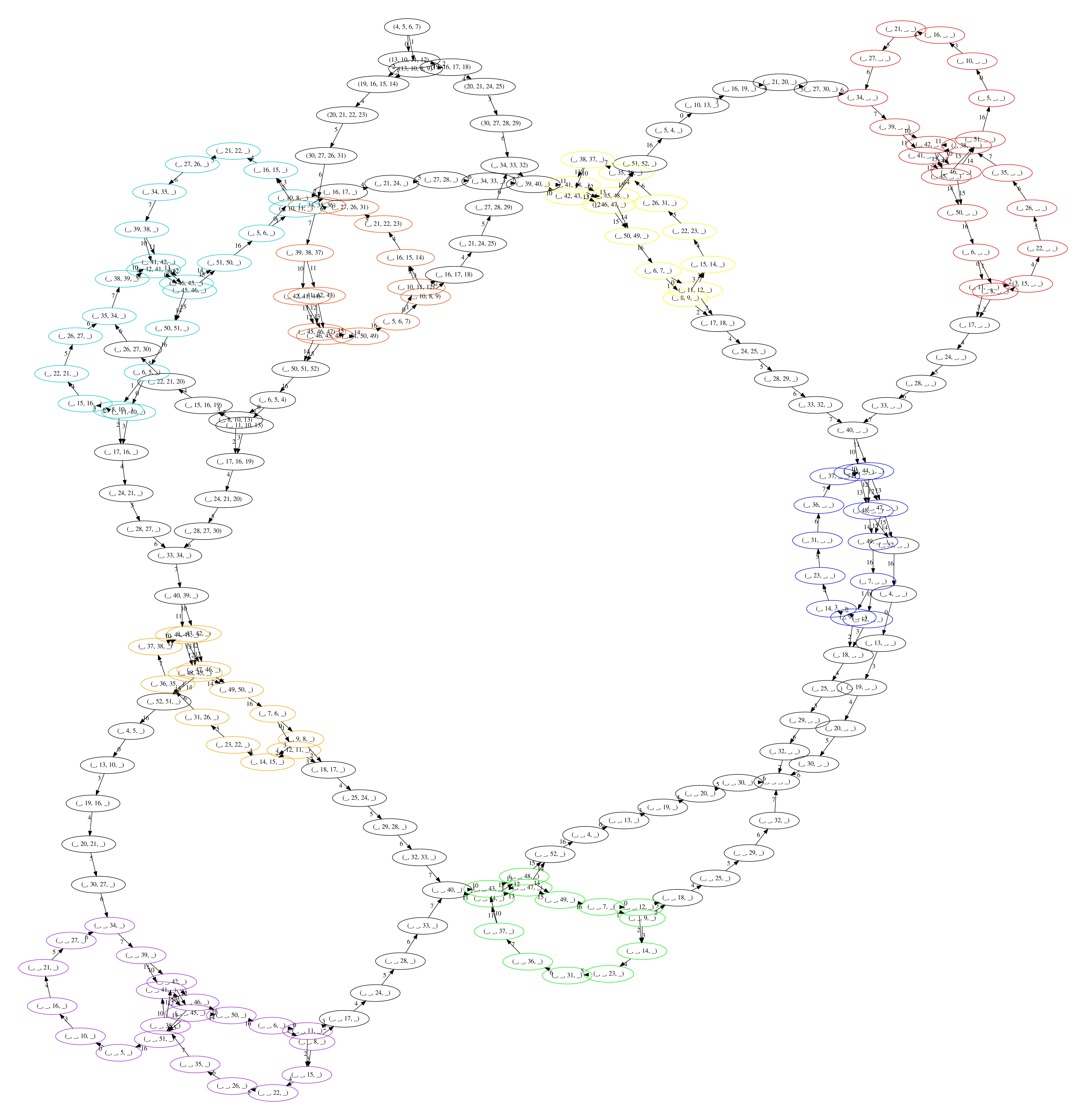}
		}
		\subfloat [] {
			  \label{fig:krebbs_cay_cycle_stereo}
			  \includegraphics[width=0.5\textwidth]{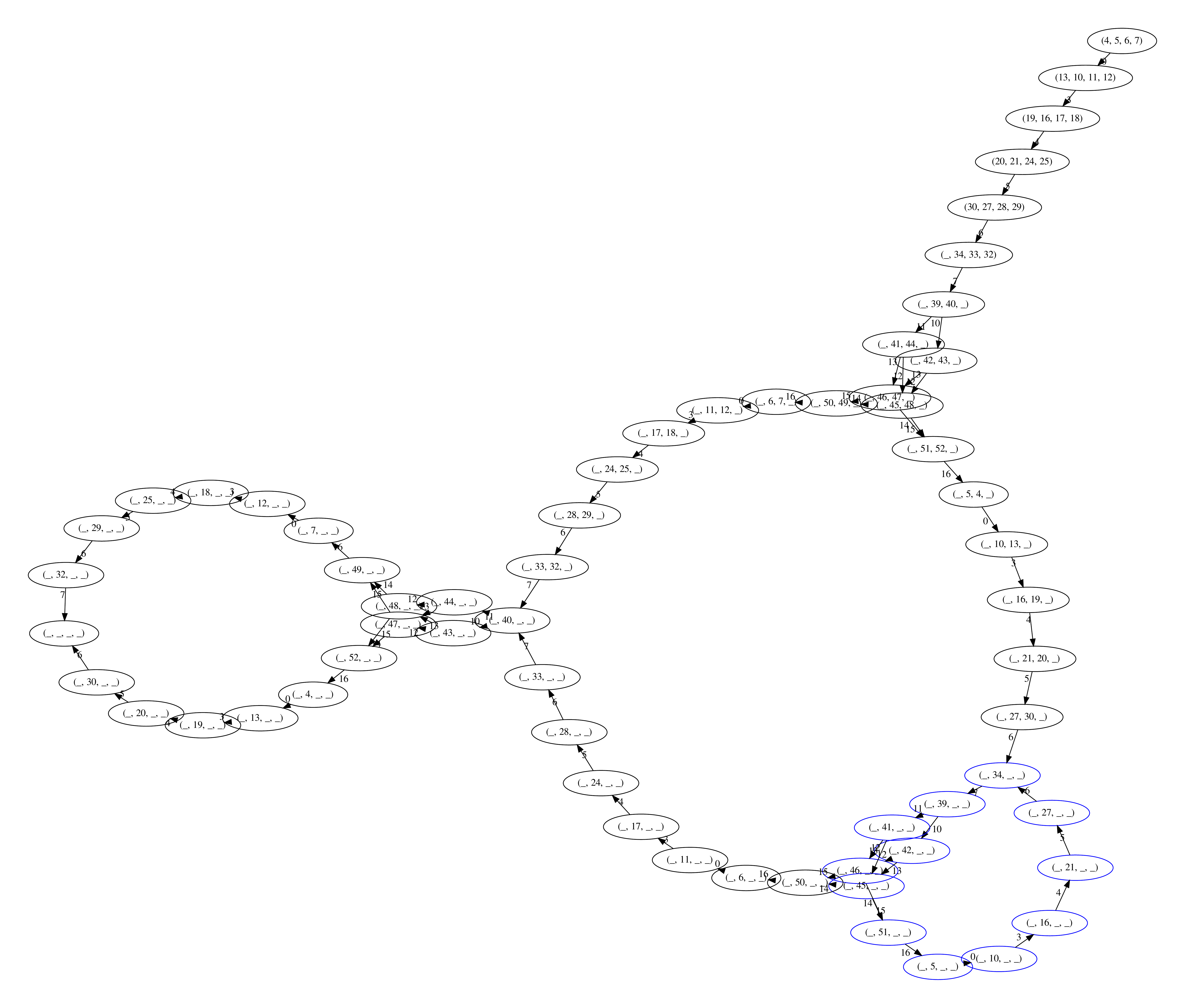}
		}
		\caption[]{
			\subref{fig:krebbs_cay_full} 
			The projected Cayley graph $\pcay(\CN, (4,5,6,7))$ wrt. $\TCA$. 
			The non black colored vertices of the same color, correspond to atom states
			that are part of the same strongly connected component.
			\subref{fig:krebbs_cay_cycle_stereo} 
			The projected cayley graph $\pcay(\CN, (4,5,6,7)$ wrt.  $\sTCA$.
			The non black colored vertices of the same color, correspond to atom states
			that are part of the same strongly connected component.
		}
	\end{figure}

	To start the cycle, an Acetyl-CoA molecule is condensed
	with an OAA molecule, executing a cycle of reactions
	that ends up regenerating the OAA molecule while expelling two \ch{CO2}
	and water on the way.  
	When an original atom is expelled from the cycle, we will
	consider it permanently lost. The carbon atoms of the OAA molecule
	that we are interested in tracking are annotated with the ids 4, 5, 6,
	and 7. Let $\bar{z} = (4,5,6,7)$. The projected Cayley graph of 
	$\pcay(\CN, \bar{z})$ wrt. TCA-$\square$ (resp. TCA-${\tetrahedron}$) ,
	consists of 213 (resp. 67) vertices. The full Cayley graphs
	are depicted in \figname \ref{fig:krebbs_cay_full} and
	\ref{fig:krebbs_cay_cycle_stereo} respectively.
	When a carbon atom leaves the TCA cycle we denote it by $"\_"$. E.g. the
	atom state $(\_,7,6,\_)$ should be read as the original carbon atoms
	with ids $4$ and $7$ has been expelled, while the carbon atoms with
	ids $5$ and $6$ are located at the atoms with id $7$ and $6$
	respectively.

\begin{figure}[t]
	\centering
	  \begin{minipage}[b]{.20\textwidth}
	  \subfloat [] {
	  \label{fig:tca_oaa_graph}\includegraphics[width=0.7\linewidth]{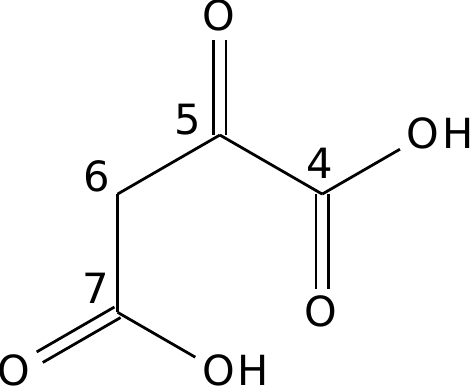}}
	\vfill
	\subfloat
	  []
	  {\label{fig:krebbs_stereo_scay}\includegraphics[width=1.2\linewidth]{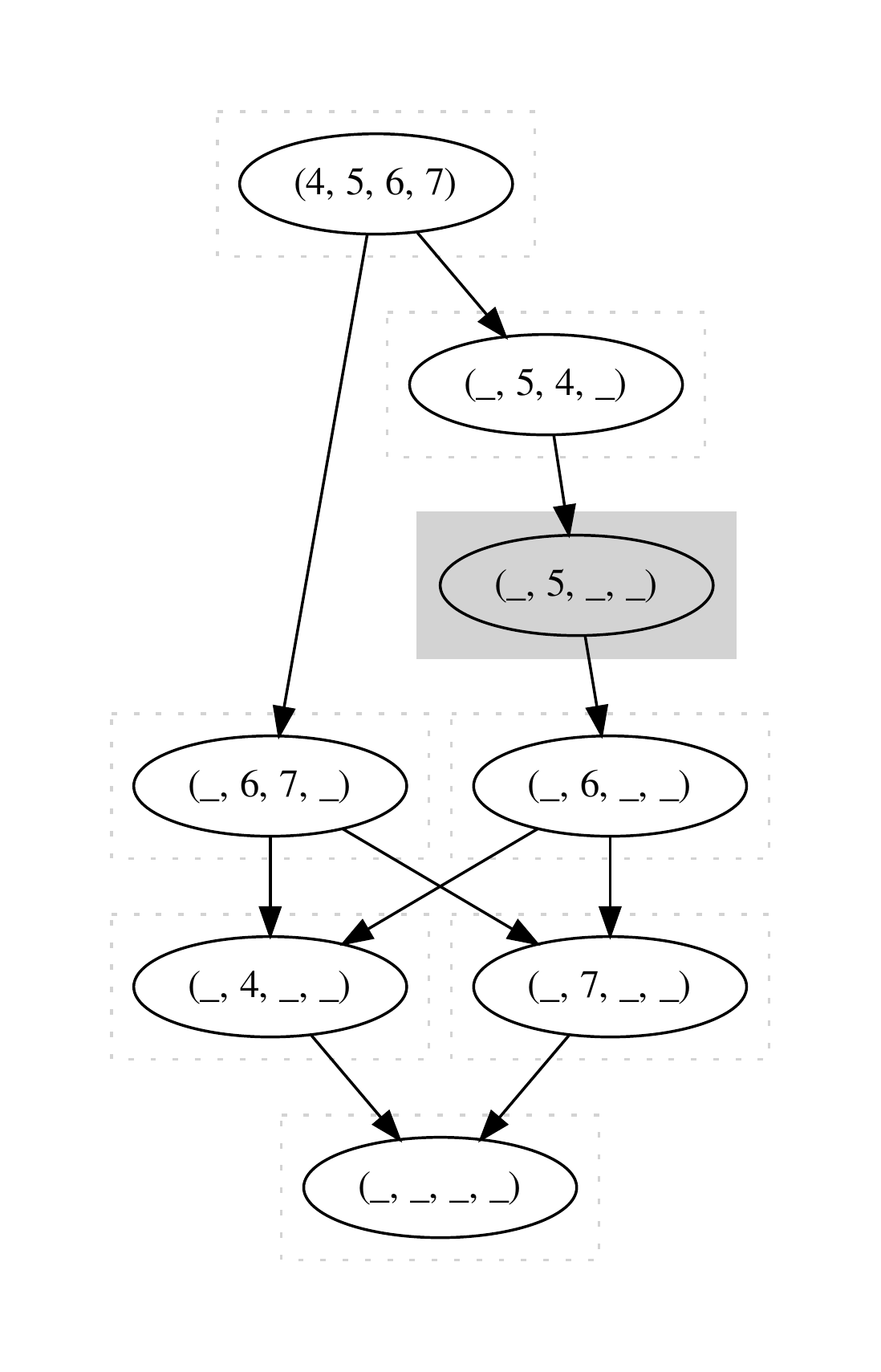}}
	  \end{minipage}
	\begin{minipage}[b]{.45\textwidth}
	\centering
	\subfloat []
	{\label{fig:krebbs_cay_simple}\includegraphics[width=1.1\linewidth]{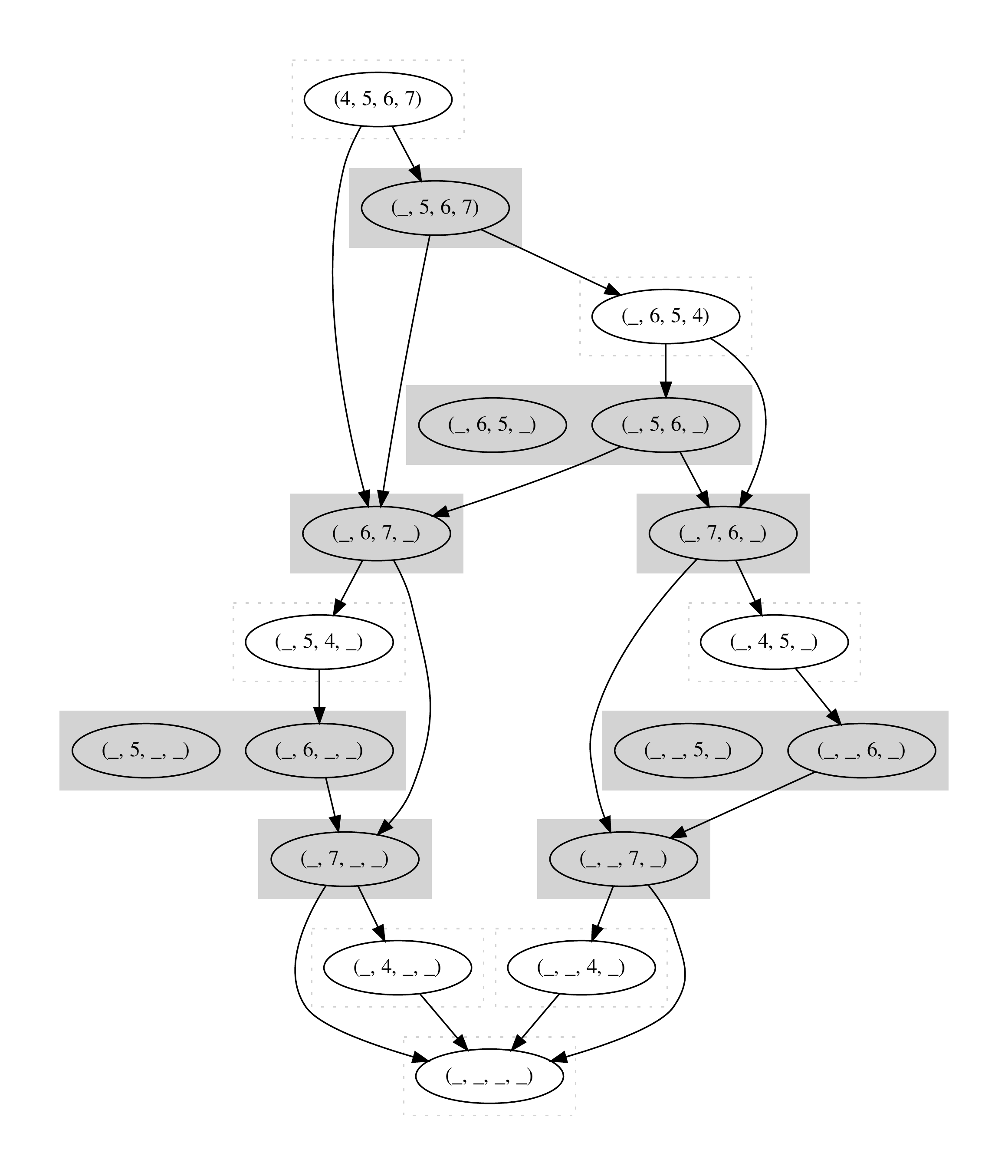}}
	\end{minipage}
	\caption[]{\subref{fig:tca_oaa_graph} The oxaloacetate molecule.
			The carbon atoms are equipped with the ids 4, 5, 6, and
			7.
			\subref{fig:krebbs_stereo_scay} The simplified projected
			Cayley graph $\scay(\CN, (4,5,6,7))$, when adjusting for
			stereospecific citrate in $tr(\CN)$.
			\subref{fig:krebbs_cay_simple} The simplified
			projected Cayley graph $\scay(\CN, (4,5,6,7))$ when not
		  	considering stereospecificity.
	}
	\label{fig:vier}
	\end{figure}

	We can find the natural subsystems of $\CN$ as the strongly connected
	components of $\pcay(\CN, \bar{z})$. In TCA-$\square$
	(resp. TCA-${\tetrahedron}$) we find 92 (resp. 51) strongly connected
	components of which 8 (resp. only 1) are non-trivial.  Any
	non-trivial strongly
	connected component must invariably contain at least one tour around
	the TCA cycle, since this is the only way the original atoms of the
	OAA molecules can be reused to create another OAA molecule. Moreover,
	any non-trivial strongly connected component represents a sequence(s)
	of reactions that uses (some of the) original atoms of the OAA molecule.
	To simplify
	$\pcay(\CN, \bar{z})$ such that only the information on carbon traces
	of the atoms of OAA are depicted, we will construct the simplified
	projected Cayley graph, denoted $\scay(\CN, \bar{z})$, as follows:
	collapse any vertex in $\pcay(\CN, \bar{z})$ that is part of a trivial
	strongly connected component and whose atoms are not located in an OAA
	molecule. Moreover, for any non-trivial strongly connected component,
	hide the edges between atom states in the same strongly connected
	component, and finally only include atom states if the atoms are
	located in a OAA molecule.  The resulting graphs for TCA-$\square$ and
	TCA-${\tetrahedron}$ are depicted in
	\figname \ref{fig:vier}. Each box in the figure represents a
	natural subsystem that contains an atom state where every atom is
	either expelled or located in an OAA molecule. When ignoring the
	stereochemical formation of citrate, $(\_,5,6,7)$ is a grey node in
	$\scay(\CN, \bar{z})$ (i.e., a representative of a strongly connected
	component $\pcay(\CN, \bar{z})$), i.e., there is a trajectory where
	three of the four original carbons of OAA are re-used at the same
	location after a TCA-$\square$ cycle
	turnover. However in TCA-$\tetrahedron$ only $(\_,5,\_,\_)$ is a
	representative of a strongly connected component, i.e., only the
	carbon with id 5 of OAA can be kept at the same location when a multitude of
	TCA-$\tetrahedron$ turnovers are executed. If that carbon changes
	location it will leave the TCA cycle after exactly two more turnovers
	(the natural subsystems reachable from $(\_,5,\_,\_)$ do not
	correspond to strongly connected components) via
	positions $5 \rightarrow 6 \rightarrow 4 \rightarrow \_$ or via 
	$5 \rightarrow 6 \rightarrow 7 \rightarrow \_$. To the best of our
	knowledge such investigations have not been executed formally before.

	Interestingly, $\scay(\CN, \bar{z})$, as depicted in \figname
	\ref{fig:krebbs_cay_simple}, allows us to closely examine each of the possible 
	carbon trajectories of TCA$-\square$.
	E.g. the fact that the atom
	state $(\_,6,7,\_)$ is present in $\scay(\CN, \bar{z})$ wrt.
	$\TCA$,
	means that there exists a sequence of reactions that expels the
	carbons with ids $4$ and $7$, but re-uses the carbon atoms with id $5$
	and $6$ to create a new OAA atom, where $5$ is located at $6$ and $6$
	is located at $7$. Structurally the atoms $4$ and $7$ corresponds to
	the the outer atoms in the carbon backbone in the OAA molecule, while
	the atoms $5$ and $6$ correspond to the inner atoms in the carbon
	backbone. In other words the presence of $(\_,6,7,\_)$, means
	there exists a sequence of reactions that expels the outer atoms of
	the carbon backbone while recycling the inner atoms.

	\figname \ref{fig:krebbs_cay_simple}, gives us a rough road map to
	determine exactly
	what sequence of events must have taken place in order to end up in
	the atom state $(\_,6,7,\_)$.  
	We start with the atom state $(4,5,6,7)$ and see there is an
	edge directly to $(\_,6,7,\_)$, meaning that we can expel the two
	outer atoms in a single cycle. This is, however, not the only way
	we can end up with the atom state $(\_,6,7,\_)$. E.g. after one
	cycle we can expel the carbon with id $4$ and end up with the atom
	state $(\_, 5, 6, 7)$, i.e., all other atoms are still in their
	original positions. After another cycle we can end up in the atom
	state $(\_,6,7,\_)$ or $(\_6,5,4)$. Note, that $(\_, 5, 6, 7)$
	is part of a non-trivial strongly connected component, meaning that
	there exists a sequence of reactions in the TCA cycle that ends up
	in the exact same atom state. i.e., we expel the carbon atom at
	position $4$ (which is already expelled) while keeping all other atoms
	at their original position. In contrast, the atom state $(\_,
	6,5,4)$ is part of a trivial strongly connected component, meaning that any
	sequence of reaction in the TCA cycle will have to change the atom
	state.

\begin{figure}
			\centering
		  \includegraphics[width=0.7\textwidth]{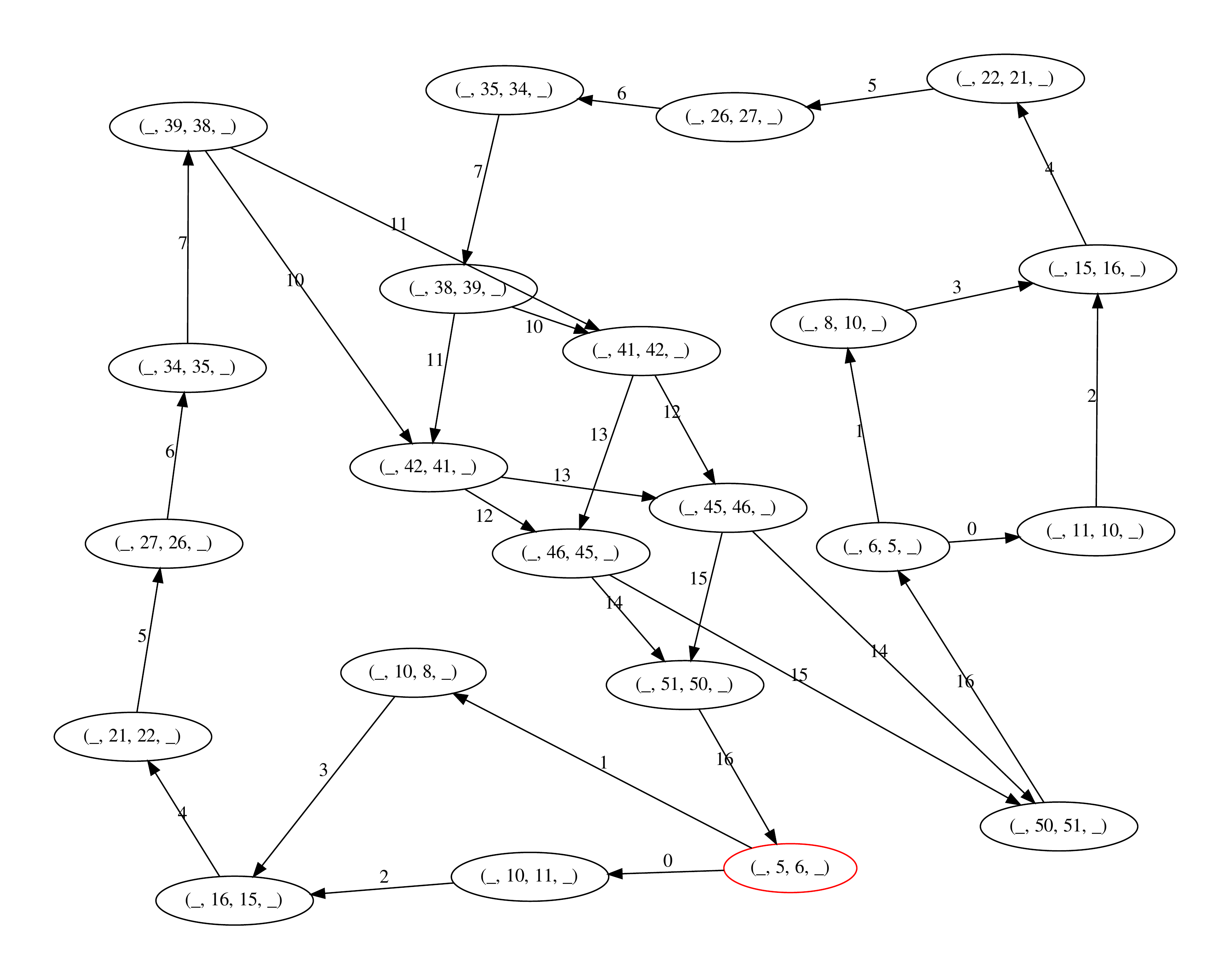}
		  \caption{
		  The strongly connected
		  component of $\pcay(\CN, (4,5,6,7))$ wrt. $\TCA$ containing the
		  state $(\_,6,5,\_)$ and $(\_,5,6,\_)$.
		  }
		  \label{fig:krebbs_cay_cycle}
	\end{figure}

	If any non-trivial strongly connected component in \figname
	\ref{fig:krebbs_cay_simple} contains more than one vertex, it means
	that we can swap between atom states after a tour in the TCA
	cycle. As an example, consider the atom state $(\_,6,5,\_)$
	and $(\_,5,6,\_)$ that are both part of the same strongly connected
	component. The fact that they are part of the same strongly connected
	component, means it is possible to swap the inner atoms of the carbon
	backbone during a TCA cycle. If we would be interested in the exact
	sequence of transformations that lead to the swap, we simply examine the subgraph of
	$\pcay(\CN, \bar{z})$ wrt. $\TCA$ corresponding to that natural subsystem of
	$\scay(\CN, \bar{z})$ wrt. $\TCA$ as illustrated in \figname 
	\ref{fig:krebbs_cay_cycle}. The figure depicts all possible ways to
	swap the positions of atoms with ids $5$ and $6$ as the possible
	paths between $(\_,5,6,\_)$ and $(\_,6,5,\_)$.
	\figname \ref{fig:krebbs_cay_path}, shows one such path traversing the
	TCA cycle without expelling any of the remaining carbon atoms.

	\newpage
	\section{Conclusion}
	In this work we have extended the insights provided by
	\cite{IsotopeLabel2019}, by showing the natural
	relationship between event traces, the characteristic monoid and
	its corresponding Cayley graph.  
	The projected Cayley graph provides valuable
	insights into local substructures of reversible event traces. 

	We see future steps for this approach to branch in at least two
	directions. On one hand, these methods shows
	obvious applications in isotopic labeling design. To this end, it
	is natural to extend the system to model the actual process of
	such experiments. E.g. when doing isotopic labeling experiments
	with mass spectrometry, molecules are broken into fragments and
	the weight of such fragments are deduced to determine the
	topology of the fragment. Using our model to track where the atoms
	might end up in such fragments and how it affects their weight seems like a natural next step.  
	On the other hand, a more rigorous investigation of the
	fundamental properties derived from semigroup theory of the
	characteristic monoid seems
	appealing. As we have shown here, understanding such relations
	might grant insights into the nature of the examined system.

    \newpage
    \section*{Acknowledgements}
This work is supported in by Novo Nordisk
Foundation grant NNF19OC0057834 and by the Independent Research Fund
Denmark, Natural Sciences, grant DFF-0135-00420B.
    
 \section*{Author Disclosure Statement}
    Nothing to declare.

    \newpage
    \singlespacing
    \printbibliography

    \end{document}